\documentclass[preprintnumbers,amsmath,amssymb]{revtex4}
\usepackage{graphicx}
\usepackage{amsmath}\usepackage{slashed}
\usepackage{color}
\usepackage{mathtools}
\usepackage{txfonts}

\UseRawInputEncoding
\begin{document}

\thispagestyle{empty}

\title{Temperature Dependence of the Response Functions of Graphene:
Impact on Casimir and Casimir-Polder Forces in and out of Thermal Equilibrium}

\author{
G.~L.~Klimchitskaya}
\affiliation{Central Astronomical Observatory at Pulkovo of the Russian Academy of Sciences, St.Petersburg,
196140, Russia}
\affiliation{Peter the Great Saint Petersburg
Polytechnic University, Saint Petersburg, 195251, Russia}

\author{
V.~M.~Mostepanenko}
\affiliation{Central Astronomical Observatory at Pulkovo of the Russian Academy of Sciences, St.Petersburg,
196140, Russia}
\affiliation{Peter the Great Saint Petersburg
Polytechnic University, Saint Petersburg, 195251, Russia}

\begin{abstract}
We review and obtain some new results on the temperature dependence
of spatially nonlocal response functions of graphene and their applications to
calculation of both the equilibrium and nonequilibrium Casimir and Casimir-Polder
forces. After a brief summary of the properties of the polarization tensor of
graphene obtained within Dirac model in the framework of quantum field theory,
we derive the expressions for the longitudinal and transverse dielectric functions.
The behavior of these functions at different temperatures is investigated in the
regions below and above the threshold. Special attention is paid to the double
pole at zero frequency which is present in the transverse response function of
graphene. An application of the response functions of graphene to calculation of
the equilibrium Casimir force between two graphene sheets and Casimir-Polder
forces between an atom (nanoparticle) and a graphene sheet is considered with due
attention to the role of a nonzero energy gap, chemical potential and a material
substrate underlying the graphene sheet. The same subject is discussed for
out-of-thermal-equilibrium Casimir and Casimir-Polder forces. The role of the
obtained and presented results for fundamental science and nanotechnology
is outlined.
\end{abstract}

\maketitle

\section{Introduction}

It has been known that the Casimir and Casimir-Polder forces act between two parallel
plates and a microparticle and a plate, respectively. These forces are caused by
fluctuations of the electromagnetic field whose spectrum is altered due to the
boundary conditions imposed on the surfaces of interacting bodies \cite{1,2}.
By now there is considerable literature devoted to the Casimir and Casimir-Polder
forces, as well as to their applications in different fields of fundamental and
applied physics (see, e.g., monographs \cite{3,4,5,6} and references therein).
The general theory of van der Waals, Casimir and Casimir-Polder forces, which are
also called dispersion forces, was created by Lifshitz \cite{7,8,9}. In this
theory, the force is expressed as a functional of the frequency-dependent dielectric
functions of plate materials and the dynamic polarizabilities of microparticles.

The original Lifshitz theory was formulated for the bodies in the state of thermal
equilibrium with the environment at some temperature $T$. In doing so, the obtained
forces depend on temperature. For dielectric plates, whose response functions to
the action of the electromagnetic field are temperature-independent, the force
dependence on the temperature is completely determined by a summation over the
Matsubara frequencies in the Lifshitz formula. It is common knowledge that the
response functions of metals depend on temperature through the relaxation parameter.
Calculations show, however, that in the state of thermal equilibrium this
dependence makes only a minor impact on the force value \cite{10,11}. As a result,
for metallic test bodies the temperature dependence of the Casimir and Casimir-Polder
forces is also  mostly determined by a summation over the Matsubara frequencies.

With an advent of two-dimensional materials, of which the most popular is graphene,
the problem of temperature-dependence of dispersion forces is taking new features.
The point is that the massless or very light quasiparticles in graphene are described
by the (2+1)-dimensional Dirac equation where the speed of light $c$ is replaced with
Fermi velocity $v_F\approx c/300$  \cite{12,13,13a,13b,13c,13d,14}. As a result, in addition
to the traditional effective temperature $T_{\rm eff}=\hbar c/(2ak_B)$, where $k_B$ is the
Boltzmann constant and $a$ is the separation distance between the Casimir plates,
there appears one more  temperature parameter $T_{\rm eff}^g=\hbar v_F/(2ak_B)$.
At $a=1~\mu$m, one has $T_{\rm eff}\approx 1145~$K but $T_{\rm eff}^g\approx 3.82~$K.

Consequently, as it was first proven in \cite{15}, for graphene the thermal regime of
the Casimir force starts at much shorter separations than for conventional 3D materials.
What is more, the response functions of graphene to the action of the electromagnetic
field are substantially temperature-dependent. Hence, the dependence of the Casimir and
Casimir-Polder forces in graphene systems on temperature at the moderate experimental
separations is equally contributed by the Matsubara summation and by the explicit
dependence on $T$ of the response functions of graphene \cite{16}. At a later time,
several other two-dimensional materials were created, such as silicene \cite{ 16.1, 16.2,16.3},
germanene \cite{16.4,16.5,16.6}, stanene \cite{16.7,16.8,16.9},
phosphorene \cite{16.10,16.11,16.12}, etc.

Many different approaches have been used in the literature for theoretical description
of the electromagnetic response of graphene in terms of the electric conductivity,
dielectric functions, density-density correlation functions, etc. Among them there are
the hydrodynamic model, the two-dimensional Drude model, Boltzmann's transport theory,
modeling in the random phase approximation and others (see articles
\cite{17,18,19,20,21,22,23,24,25,26,27,28,29,30,31,32,33,34,35,36,37,38,39,40,41,42,43,44,44a,45,46,47,48,49,50,51,52,53,54,55,55a}
and reviews \cite{56,57,58}).

The fundamental difference between the response functions of graphene and the regular
3D materials is that in the application region of the Dirac model, i.e., at energies
below 3~eV, the former can be found on the basis of first principles of thermal
quantum field theory by calculating the loop diagram of electronic quasiparticles with
two photon legs. This diagram represents the polarization tensor of graphene calculated
at both zero and nonzero temperature using the methods of standard and thermal quantum
field theory, respectively \cite{59,60,61,62,63,64,65}. The polarization tensor of
graphene depends on the frequency $\omega$, the two-dimensional wave vector
$\mbox{\boldmath$k$}=(k^1,k^2)$, and the temperature. For a graphene with a nonzero
mass of quasiparticles $m$ it also depends on the energy gap parameter $\Delta=2mv_F^2$
and for a graphene doped with foreign atoms other than C --- on the chemical potential
$\mu$ \cite{12,13,14,56,57,58}.

In \cite{65} the polarization tensor of graphene depending on all these parameters was
found at only the discrete Matsubara frequencies $\omega=i\xi_l=2\pi ik_BTl/\hbar$,
where $l=0,\,1,\,2,\,\ldots\,$. The correct analytic continuation of the obtained
expressions to the entire plane of complex frequencies, including the real frequency axis,
was obtained for a gapped but undoped graphene in \cite{66} and, for a doped graphene,
in \cite{67}. The spatially nonlocal tensor of electric conductivity and the dielectric
tensor of graphene are immediately expressed via the polarization tensor \cite{68}.
This opens opportunities for a computation of the temperature-dependent Casimir and
Casimir-Polder forces in graphene systems both in thermal equilibrium and in situations
when the state of thermal equilibrium is violated.

In this review, which also contains new results (see Sections~3 and 4), we discuss the
temperature dependence of the spatially nonlocal longitudinal and transverse dielectric
functions of graphene expressed via the polarization tensor. Although the general
expressions for these quantities are available in the literature and used in computations,
the analysis of their temperature dependence is still lacking. Then we consider the thermal
effects in the Casimir and Casimir-Polder forces in graphene systems in the state of
thermal equilibrium and when the condition of thermal equilibrium is violated. Special
attention is focused on the classical limit of Casimir and Casimir-Polder forces.

This review is organized as follows. In Section~2, we consider the polarization,
conductivity and dielectric tensors
of graphene, their interrelation, and different representations for the reflection
coefficients on a graphene sheet. Section~3 is devoted to the temperature dependence
of the longitudinal and transverse dielectric functions of graphene at frequencies
below the threshold $\omega=v_F|\mbox{\boldmath$k$}|$. The temperature dependence
of these functions at frequencies above the threshold is analyzed in Section~4.
Thermal effects in the Casimir force between two graphene sheets, both freestanding
and deposited on a substrate, are reviewed in Section~5. Section~6 contains the
discussion of the same subject for the case of the Casimir-Polder force. Thermal
effects in the Casimir and Casimir-Polder forces in situations out of thermal
equilibrium are considered in Section~7. Finally, Sections~8 and 9 are devoted to
the discussion and our conclusions, respectively. The Gaussian system of units is used.

\section{Main Quantities: Polarization Tensor, Electric Conductivity,
Dielectric Functions, and Reflection Coefficients}
\newcommand{\wkt}{{(\omega,\mbox{\boldmath$k$},T)}}
\newcommand{\lx}{{\chi^{\rm L}}}
\newcommand{\tx}{{\chi^{\rm T}}}
\newcommand{\ve}{{\varepsilon}}
\newcommand{\lve}{{\varepsilon^{\rm L}}}
\newcommand{\tve}{{\varepsilon^{\rm T}}}
\newcommand{\vkw}{{\sqrt{v_F^2k^2-\omega^2}}}
\newcommand{\wvk}{{\sqrt{\omega^2-v_F^2k^2}}}
\newcommand{\qt}{{\tilde{q}}}
The polarization tensor of graphene
$\Pi_{\mu\nu}\wkt$ with $\mu,\,\nu=0,\,1,\,2$
represents the Feynman diagram consisting of an electronic quasiparticle loop
with two photon legs.  {We define the polarization tensor as in} \cite{66},
 {but here do
not put $\hbar=c=1$. The definition of} \cite{66}
 { exploits the metrical tensor
$g_{\mu\nu}={\rm diag}\{1,-1,-1\}$, the Feynman propagators, and the two-sided Fourier
transforms.}
Due to the gauge invariance, the polarization tensor satisfies
the transversality condition \cite{59,60,61,62,63,64,65,66,67}
\begin{equation}
k^{\mu}\Pi_{\mu\nu}(\omega,\mbox{\boldmath$k$},T)=0.
\label{eq1}
\end{equation}
\noindent
In the absence of constant  {in time, external} magnetic field, the polarization tensor is symmetric,
$\Pi_{\mu\nu}=\Pi_{\nu\mu}$, and all its components can be expressed in terms of two
\cite{65}. It is convenient to express the components of $\Pi_{\mu\nu}$ via
$\Pi_{00}$ and the following combination
\begin{equation}
\Pi(\omega,\mbox{\boldmath$k$},T)\equiv k^2\Pi_{\nu}^{\,\nu}(\omega,\mbox{\boldmath$k$},T)
+\left(\frac{\omega^2}{c^2}-k^2\right)\Pi_{00}(\omega,\mbox{\boldmath$k$},T),
\label{eq2}
\end{equation}
\noindent
where $k=|\mbox{\boldmath$k$}|=(k_1^2+k_2^2)^{1/2}$ and $\Pi_{\nu}^{\,\nu}$
with a summation over $\nu=0,\,1,\,2$ is the trace of the polarization tensor.

As noticed in Section 1, in the general case the polarization tensor also depends on the
energy gap parameter $\Delta$ and the chemical potential $\mu$ of the graphene sample.
Below, however, for the sake of brevity and simplicity of presentation, we present
the mathematical expressions for the case of pristine graphene with $\Delta=\mu=0$.
In so doing, the impact of nonzero $\Delta$ and $\mu$ on the results obtained  will
be especially indicated.

The polarization tensor of graphene is characterized by the so-called threshold
occurring at $\omega=v_Fk$. As  a result, it is convenient to present the separate
expressions for $\Pi_{00}$ and $\Pi$ in the region $0<\omega<v_Fk$ (the strongly
evanescent waves) and for $\omega>v_Fk$ (the plasmonic region of evanescent waves,
$v_Fk<\omega<ck$ \cite{69,69a,69b,69c}, and the propagating waves, $\omega\geqslant ck$).

We begin with the region $0<\omega<v_Fk$. In this region the real part of $\Pi_{00}$
takes the form \cite{70}
\begin{eqnarray}
&&
{\rm Re}\Pi_{00}\wkt=\frac{\pi e^2k^2}{\vkw}+\frac{8e^2}{v_F^2}\left\{
\vphantom{\left[\int\limits_{0}^{v_Fk}\right]}
\frac{2k_BT \ln 2}{\hbar}+\frac{1}{2\vkw}\right.
\nonumber \\
&&~~~~~
\times \left.\left[\int\limits_{0}^{v_Fk-\omega}\!\!\!dx w(x,T)f_1(x)-
\int\limits_{0}^{v_Fk+\omega}\!\!\!dx w(x,T)f_2(x)
\right]\right\},
\label{eq3}
\end{eqnarray}
\noindent
where $e$ is the electron charge and
\begin{equation}
w(x,T)=\left[\exp\left(\frac{\hbar x}{2k_BT}\right)+1\right]^{-1},
\qquad
f_{1,2}(x)=\left[v_F^2k^2-(x\pm\omega)^2\right]^{1/2}.
\label{eq4}
\end{equation}

In a similar way, for the imaginary part of $\Pi_{00}$ one obtains \cite{70}
\begin{equation}
{\rm Im}\Pi_{00}\wkt=\frac{4e^2}{v_F^2\vkw}\left[
\int\limits_{v_Fk-\omega}^{\infty}\!\!\!dx w(x,T)f_3(x)-
\int\limits_{v_Fk+\omega}^{\infty}\!\!\!dx w(x,T)f_4(x)
\right],
\label{eq5}
\end{equation}
\noindent
where
\begin{equation}
f_{3,4}(x)=\left[(x\pm\omega)^2-v_F^2k^2\right]^{1/2}.
\label{eq6}
\end{equation}

In the same region, the real part of $\Pi$ is given by \cite{70}
\begin{eqnarray}
&&
{\rm Re}\Pi\wkt=\frac{\pi e^2k^2}{c^2}\vkw+\frac{8e^2}{v_F^2c^2}\left\{
\vphantom{\left[\int\limits_{0}^{v_Fk}\right]}
\frac{2\omega^2k_BT \ln 2}{\hbar}+\frac{1}{2}\vkw\right.
\nonumber \\
&&~~~~~
\times \left.\left[\int\limits_{0}^{v_Fk-\omega}\!\!\!dx w(x,T)
\frac{(x+\omega)^2}{f_1(x)}-
\int\limits_{0}^{v_Fk+\omega}\!\!\!dx w(x,T)\frac{(x-\omega)^2}{f_2(x)}
\right]\right\}.
\label{eq7}
\end{eqnarray}

Finally, for the ${\rm Im}\Pi$, the following result occurs \cite{70}
\begin{equation}
{\rm Im}\Pi\wkt=\frac{4e^2}{v_F^2c^2}\vkw\left[
\int\limits_{v_Fk+\omega}^{\infty}\!\!\!dx w(x,T)
\frac{(x-\omega)^2}{f_4(x)}-
\int\limits_{v_Fk-\omega}^{\infty}\!\!\!dx w(x,T)
\frac{(x+\omega)^2}{f_3(x)}
\right].
\label{eq8}
\end{equation}

Now we deal with the remaining region $\omega>v_Fk$. In this region, the real and
imaginary parts of $\Pi_{00}$ are presented as
\begin{eqnarray}
&&
{\rm Re}\Pi_{00}\wkt=\frac{4e^2}{v_F^2}\left\{
\vphantom{\left[\int\limits_{0}^{v_Fk}\right]}
\frac{4k_BT \ln 2}{\hbar}-\frac{1}{\wvk}\left[
\int\limits_{0}^{\infty}dxw(x,T)f_3(x)\right.\right.
\nonumber \\
&&~~~~~
\left.\left.-\int\limits_{v_Fk+\omega}^{\infty}\!\!\!dx w(x,T)f_4(x)+
\int\limits_{0}^{v_Fk-\omega}\!\!\!dx w(x,T)f_4(x)
\right]\right\}
\label{eq9}
\end{eqnarray}
\noindent
and
\begin{equation}
{\rm Im}\Pi_{00}\wkt=\frac{e^2}{\wvk}\left[\pi k^2-\frac{4}{v_F^2}
\int\limits_{-v_Fk}^{v_Fk}\!\!\!dx w(\omega+x,T)
\sqrt{v_F^2k^2-x^2}\right].
\label{eq10}
\end{equation}

For $\Pi$ in the region $\omega>v_Fk$ one finds \cite{70}
\begin{eqnarray}
&&
{\rm Re}\Pi\wkt=\frac{4e^2}{v_F^2c^2}\left\{
\vphantom{\left[\int\limits_{0}^{v_Fk}\right]}
\frac{4\omega^2k_BT \ln 2}{\hbar}-\vkw\left[
\int\limits_{0}^{\infty}dxw(x,T)\frac{(x+\omega)^2}{f_3(x)}
\right.\right.
\nonumber \\
&&~~~~~
\left.\left.-\int\limits_{v_Fk+\omega}^{\infty}\!\!\!dx w(x,T)
\frac{(x-\omega)^2}{f_4(x)}+
\int\limits_{0}^{v_Fk-\omega}\!\!\!dx w(x,T)\frac{(x-\omega)^2}{f_4(x)}
\right]\right\}
\label{eq11}
\end{eqnarray}
\noindent
and
\begin{equation}
{\rm Im}\Pi\wkt=\frac{e^2}{v_F^2c^2}\vkw\left[-\pi v_F^2k^2+
4\int\limits_{-v_Fk}^{v_Fk}\!\!\!dx w(\omega+x,T)
\frac{x^2}{\sqrt{v_F^2k^2-x^2}}
\right].
\label{eq12}
\end{equation}

The polarization tensor in (\ref{eq3}), (\ref{eq5}), (\ref{eq7})--(\ref{eq12})
essentially depends on the wave vector $\mbox{\boldmath$k$}$.
Because of this the response of
graphene to the electromagnetic field is spatially nonlocal. In terms of the
polarization tensor, the tensor of electric conductivity is expressed as
\cite{24,71,73}
\begin{equation}
\sigma^{\mu\nu}\wkt=\frac{c^2}{4\pi\hbar}\,\frac{\Pi^{\mu\nu}\wkt}{i\omega}.
\label{eq13}
\end{equation}

Similar to the polarization tensor, in the absence of a constant  {in time, external} magnetic field
the tensor of electric conductivity has two independent components. It is common
to characterize it by the longitudinal and transverse conductivities \cite{74},
which are expressed via the polarization tensor as \cite{75,76,77,78}
\begin{equation}
\sigma^{\rm L}\wkt=-\frac{i\omega}{4\pi\hbar k^2}\Pi_{00}\wkt
\label{eq14}
\end{equation}
\noindent
and
\begin{equation}
\sigma^{\rm T}\wkt=\frac{ic^2}{4\pi\hbar k^2\omega}\Pi\wkt.
\label{eq15}
\end{equation}

Note also that the longitudinal and transverce conductivities are closely related to the
longitudinal and transverse electric susceptibilities and, thus, corresponding dielectric
functions. For the two-dimensional materials, this relation takes the form \cite{6,47}
\begin{equation}
\chi^{\rm L,T}\wkt=\ve^{\rm L,T}\wkt-1=\frac{2\pi ik}{\omega}\sigma^{\rm L,T}\wkt.
\label{eq16}
\end{equation}

{}From (\ref{eq15}) and (\ref{eq16}) it is easy to express the electric susceptibilities
and dielectric functions of graphene via the polarization tensor. The results are
\begin{equation}
\lx\wkt=\lve\wkt-1=\frac{1}{2\hbar k}\Pi_{00}\wkt
\label{eq17}
\end{equation}
\noindent
and
\begin{equation}
\tx\wkt=\tve\wkt-1=-\frac{c^2}{2\hbar k\omega^2}\Pi\wkt.
\label{eq18}
\end{equation}
\noindent
 {Thus, the response of graphene to the electromagnetic field can be described on
equal terms either by $\Pi_{00}$ and $\Pi$, or by $\sigma^{{\rm L,T}}$, or by $\varepsilon^{{\rm L,T}}$.}

It is common knowledge that the present-day formulation of the Lifshitz theory expresses
the Casimir and Casimir-Polder forces via the reflection coefficients on the interacting
surfaces \cite{5,6}. Using the standard electrodynamic
boundary conditions, the reflection coefficients
on the graphene sheet were expressed via the polarization tensor for two independent
polarizations of the electromagnetic field, transverse magnetic (TM) and transverse
electric (TE) \cite{64,65}
\begin{equation}
r_{\rm TM}\wkt=\frac{q\Pi_{00}\wkt}{q\Pi_{00}\wkt +2\hbar k^2}
\label{eq19}
\end{equation}
\noindent
and
\begin{equation}
r_{\rm TE}\wkt=-\frac{\Pi\wkt}{\Pi\wkt +2\hbar k^2q}
\label{eq20}
\end{equation}
\noindent
where $q=(k^2-\omega^2/c^2)^{1/2}$.

Using (\ref{eq14}) and (\ref{eq15}), these reflection coefficients can be equivalently
expressed via the longitudinal and transverse conductivities \cite{43,79}
\begin{equation}
r_{\rm TM}\wkt=\frac{2\pi iq\sigma^{\rm L}\wkt}{2\pi iq\sigma^{\rm L}\wkt+\omega}
\label{eq21}
\end{equation}
\noindent
and
\begin{equation}
r_{\rm TE}\wkt=-\frac{2\pi\omega\sigma^{\rm T}\wkt}{2\pi\omega\sigma^{\rm T}\wkt+ic^2q}.
\label{eq22}
\end{equation}

Finally, with the help of (\ref{eq17}) and (\ref{eq18}), the reflection coefficients
(\ref{eq19}) and (\ref{eq20}) can be expressed via the longitudinal and transverse
dielectric functions of graphene \cite{79}
\begin{equation}
r_{\rm TM}\wkt=\frac{q[\lve\wkt-1]}{q[\lve\wkt-1]+k}
\label{eq23}
\end{equation}
\noindent
and
\begin{equation}
r_{\rm TE}\wkt=-\frac{\omega^2[\tve\wkt-1]}{\omega^2[\tve\wkt-1]-c^2qk}.
\label{eq24}
\end{equation}

For further applications, it is important also to present the reflection coefficients
of a graphene sheet deposited on a material substrate described by the dielectric
function $\ve(\omega)$ depending only on the frequency. Here, we express them in terms
of dielectric permittivities of graphene $\ve^{\rm L,T}$ \cite{79a}
\begin{equation}
R_{\rm TM}\wkt=
\frac{k[\ve(\omega)q-\qt]+2q\qt[\lve\wkt-1]}{k[\ve(\omega)q+\qt]+2q\qt[\lve\wkt-1]}
\label{eq25}
\end{equation}
\noindent
and
\begin{equation}
R_{\rm TE}\wkt=
-\frac{2\omega^2[\tve\wkt-1]+c^2k(q-\qt)}{2\omega^2[\tve\wkt-1]-c^2k(q+\qt)},
\label{eq26}
\end{equation}
where $\qt=[k^2-\ve(\omega)\omega^2/c^2]^{1/2}$. It is seen that if $\ve(\omega)=1$
(i.e., there is no substrate) we have $\qt=q$ and, as a result, (\ref{eq25})
 and (\ref{eq26}) transform into (\ref{eq23}) and (\ref{eq24}), respectively,
 as it should be.

\section{Temperature Dependence of the Dielectric Functions of Graphene
Below the Threshold}

In this section, we consider the electric susceptibilities $\lx$, $\tx$ and the
longitudinal, $\lve$, and transverse, $\tve$, dielectric functions of graphene
versus temperature in the region $\omega<v_Fk$.

The real parts of the longitudinal electric susceptibility and dielectric function in
this region are obtained by substituting (\ref{eq3}) in (\ref{eq17})
\begin{eqnarray}
&&
{\rm Re}\lx\wkt={\rm Re}\lve\wkt-1=
\frac{\pi e^2k}{2\hbar\vkw}+\frac{8e^2k_BT\ln 2}{v_F^2\hbar^2k}
\nonumber \\
&&~~~~~
+\frac{2 e^2}{v_F^2\hbar k\vkw}
\left[\int\limits_{0}^{v_Fk-\omega}\!\!\!dx w(x,T)f_1(x)-
\int\limits_{0}^{v_Fk+\omega}\!\!\!dx w(x,T)f_2(x)
\right].
\label{eq27}
\end{eqnarray}

In a similar way, the imaginary parts of the longitudinal electric susceptibility
and dielectric function for $\omega<v_Fk$
 are obtained by substituting (\ref{eq5}) in (\ref{eq17})
\begin{eqnarray}
&&
{\rm Im}\lx\wkt={\rm Im}\lve\wkt=
\frac{2 e^2}{v_F^2\hbar k\vkw}
\nonumber \\
&&~~~~~
\times\left[
\int\limits_{v_Fk-\omega}^{\infty}\!\!\!dx w(x,T)f_3(x)-
\int\limits_{v_Fk+\omega}^{\infty}\!\!\!dx w(x,T)f_4(x)
\right].
\label{eq28}
\end{eqnarray}

{}From (\ref{eq27}) and (\ref{eq28}) it is immediately obvious that
\begin{equation}
\lim\limits_{\omega\to 0}{\rm Re}\lx\wkt=\frac{\pi e^2}{2\hbar v_F}+
\frac{8e^2 k_B T\ln 2}{v_F^2\hbar^2 k}, \qquad
\lim\limits_{\omega\to 0}{\rm Im}\lx\wkt=0.
\label{eq29}
\end{equation}
\noindent
{}From (\ref{eq28}) it is also seen that ${\rm Im}\lve>0$ in accordance with the
requirements of ther\-mo\-dy\-na\-mics \cite{74}.

\begin{figure}[t]
\vspace*{-11.cm}
\centerline{\hspace*{-1.5cm}
\includegraphics[width=7.in]{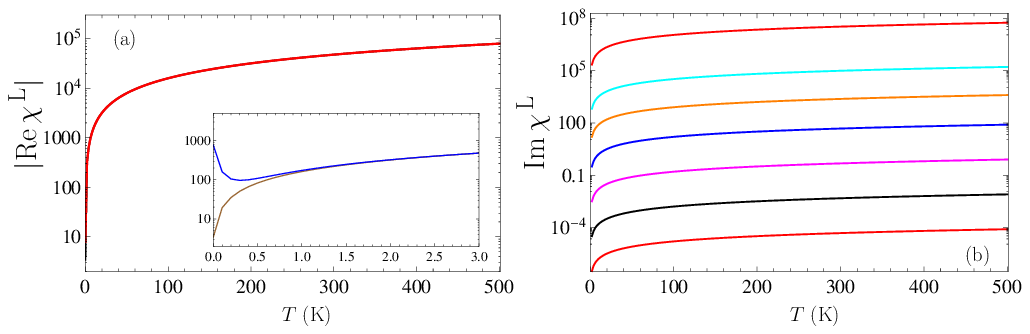}}
\vspace*{-10.2cm}
\caption{The computational results for (a) magnitude of the real part and (b)
imaginary part of the longitudinal electric susceptibility of graphene in the
region below the threshold are plotted as the functions of temperature.
$|{\rm Re}\lx|$ does not depend on $\omega$ in the wide region from
$\omega=10$ to $0.999999\times 10^{10}~$rad/s with except of the temperature
interval $0<T<1~$K (see the inset in Figure 1(a) where the bottom and top lines
are plotted for $\omega=10$ and $0.999999\times 10^{10}~$rad/s, respectively).
The lines in Figure 1(b) counted from bottom to top are plotted for
$\omega=10$, $10^3$, $10^5$, $10^7$, $5\times 10^8$, $9\times10^{10}$, and
$0.999999\times 10^{10}~$rad/s, respectively.}
\label{fg1}
\end{figure}

We computed the real and imaginary parts of the longitudinal electric susceptibility
$\lx$ by (\ref{eq27}) and (\ref{eq28}) as the functions of temperature for the fixed
wave vector $k=100~\mbox{cm}^{-1}$ and different values of $\omega$. The computational
results are presented in Figure~\ref{fg1} for (a) the magnitude of ${\rm Re}\lx$
and (b) ${\rm Im}\lx$. The line presented in Figure~\ref{fg1}(a) depends almost
not at all on the frequency  in the wide region from
$\omega=10$ to $0.999999\times 10^{10}~$rad/s (in this and below figures
$v_Fk=10^{10}~$rad/s). Some minor distinction between the lines at different
frequencies arises only at the very low temperatures. To illustrate this fact, in
the inset to Figure~\ref{fg1}(a) we plot  $|{\rm Re}\lx|$ as a function of temperature
for $\omega=10~$rad/s (bottom line) and for $\omega=0.999999\times 10^{10}~$rad/s
(top line). The lines corresponding to all intermediate frequencies are confined
between them. It is seen that some differences between the lines plotted for different
frequencies arise only at $T<1~$K.

In Figure~\ref{fg1}(b), the seven lines counted
from bottom to top are plotted for
$\omega=10$, $10^3$, $10^5$, $10^7$, $5\times 10^8$, $9\times10^{10}$, and
$0.999999\times 10^{10}~$rad/s, respectively. As is seen in Figure~\ref{fg1}(a,b), both
$|{\rm Re}\lx|$  and ${\rm Im}\lx$ increase monotonously with increasing temperature
and ${\rm Im}\lx$ also increases with increasing frequency.

Now we consider the transverse electric susceptibility and dielectric function of
graphene in the region $\omega<v_Fk$, i.e., below the threshold. By substituting
(\ref{eq7}) in (\ref{eq18}), for the real parts of these quantities one finds
\begin{eqnarray}
&&
{\rm Re}\tx\wkt={\rm Re}\tve\wkt-1=
-\frac{\pi e^2k}{2\hbar\omega^2}\vkw-\frac{8e^2k_BT\ln 2}{v_F^2\hbar^2k}
\label{eq29a} \\
&&~~
-\frac{2e^2}{v_F^2\hbar k\omega^2}\vkw
\left[\int\limits_{0}^{v_Fk-\omega}\!\!\!dx w(x,T)
\frac{(x+\omega)^2}{f_1(x)}-
\int\limits_{0}^{v_Fk+\omega}\!\!\!dx w(x,T)\frac{(x-\omega)^2}{f_2(x)}
\right].
\nonumber
\end{eqnarray}

The imaginary parts of the transverse electric susceptibility and dielectric
function are found by substituting
(\ref{eq8}) in (\ref{eq18})
\begin{eqnarray}
&&
{\rm Im}\tx\wkt={\rm Im}\tve\wkt=
\frac{2e^2}{v_F^2\hbar k\omega^2}\vkw
\nonumber \\
&&~~
\times\left[\,
\int\limits_{v_Fk-\omega}^{\infty}\!\!\!dx w(x,T)
\frac{(x+\omega)^2}{f_3(x)}-
\int\limits_{v_Fk+\omega}^{\infty}\!\!\!dx w(x,T)
\frac{(x-\omega)^2}{f_4(x)}
\right].
\label{eq30}
\end{eqnarray}

As can be seen from (\ref{eq29a}), at low frequencies satisfying the condition
$\omega\ll v_Fk$ and fixed $T\neq 0$, the difference of integrals in the square
brackets behaves as $v_F^2k^2\hbar\omega I_1/(2k_BT)$ where
\begin{equation}
I_1\equiv 2\int\limits_{0}^{1}\frac{t^2\,dt}{\sqrt{1-t^2}}\,
\frac{e^{\gamma t}}{(e^{\gamma t}+1)^2}
\label{eq31}
\end{equation}
\noindent
and $\gamma=v_Fk\hbar/(2k_BT)$. Then, for the behavior of ${\rm Re}\tx$ and
${\rm Re}\tve$ at low frequencies, one obtains from (\ref{eq29a})
\begin{equation}
{\rm Re}\tx\wkt={\rm Re}\tve\wkt-1=
-\frac{\pi e^2v_Fk^2}{2\hbar\omega^2}-\frac{8e^2k_BT\ln 2}{v_F^2\hbar^2k}
-\frac{e^2v_Fk^2}{k_BT}\,\frac{I_1}{\omega}.
\label{eq32}
\end{equation}

Along similar lines, the behavior of the difference of integrals in (\ref{eq30})
at $\omega\ll v_Fk$ is given by  $v_F^2k^2\hbar\omega I_2/(2k_BT)$ where
\begin{equation}
I_2\equiv 2\int\limits_{1}^{\infty}\frac{t^2\,dt}{\sqrt{t^2-1}}\,
\frac{e^{\gamma t}}{(e^{\gamma t}+1)^2}.
\label{eq33}
\end{equation}
\noindent
Substituting this in (\ref{eq30}), for the low-frequency behavior of
${\rm Im}\tx$ and ${\rm Im}\tve$ we find
\begin{equation}
{\rm Im}\tx\wkt={\rm Im}\tve\wkt=
\frac{e^2v_Fk^2}{k_BT}\,\frac{I_2}{\omega}.
\label{eq34}
\end{equation}
\noindent
{}From (\ref{eq30}) it follows also that ${\rm Im}\tve>0$.

As is seen from (\ref{eq32}) and (\ref{eq34}), at fixed temperature  but at low
frequencies both ${\rm Re}\tx$ and ${\rm Im}\tx$ (as well as ${\rm Re}\tve$ and
${\rm Im}\tve$) possess the simple pole at $\omega=0$ described by the last terms
on the r.h.s.{\  }of (\ref{eq32}) and (\ref{eq34}). What is more, ${\rm Re}\tx$
and ${\rm Re}\tve$ possess the double pole at $\omega=0$ described by the first term
on the r.h.s.{\  }of (\ref{eq32}).    The presence of a double pole in
the response function is an unusual feature of graphene.
It is generally believed that the response functions of dielectric materials are
regular at zero frequency whereas for metals they have the simple pole.

It has been known, however, that numerous precision experiments on measuring the
Casimir force (see \cite{5,80,81,82,83} for a review) exclude theoretical predictions
if the low-frequency behavior of metals is described by the dielectric function of
the Drude model having a simple pole at zero frequency. The experimental data of these
experiments are in good agreement with theoretical predictions using the dielectric
function of metals described by the plasma model which has the double pole at zero
frequency. This fact is considered as a puzzle because the dissipationless plasma
model should be applicable only at sufficiently high frequencies where the dissipation
processes of free charge carriers do not play any role. That is why the prediction of
the double pole at zero frequency for graphene made on the solid basis of quantum field
theory is of much interest as a signal that the commonly used semi-phenomenological
description of the response functions of 3D materials might be not complete.
What is more, measurements of the Casimir force in graphene system are in good
agreement with the theoretical predictions using the response function $\tve$
having the double pole at zero frequency \cite{84,85}.

In spite of this, it was recently claimed \cite{86} that the double pole appearing
in the transverse dielectric function of graphene is "nonphysical". In order to
remove it from (\ref{eq32}), it was suggested to replace  the polarization tensor in
(\ref{eq13}) with the modified ``regularized"  {expression} defined as
\begin{equation}
\widetilde{\Pi}^{\mu\nu}\wkt={\Pi}^{\mu\nu}\wkt-
\lim\limits_{\omega\to 0}{\Pi}^{\mu\nu}\wkt.
\label{eq35}
\end{equation}
\noindent
 {According to} \cite{86},  {equation} (\ref{eq13})  { containing the ``regularized"
expression}  (\ref{eq35})  {in place of the polarization tensor ${\Pi}^{\mu\nu}$ is obtained
by a derivation from the Kubo formula. In this derivation, however, the nonrelativistic
concept of causality was used represented by the one-sided Fourier
transforms. This is inappropriate for graphene described by the relativistic Dirac
model. If the relativistic causality realized in the form of two-sided Fourier
transforms is employed in derivation, the Kubo formula leads to equation}
(\ref{eq13})   {with the correct polarization tensor ${\Pi}^{\mu\nu}$.}
It was shown that in the framework of quantum field theory the
polarization tensor ${\Pi}^{\mu\nu}$ is defined uniquely and its modification would
result in violation of fundamental physical principles \cite{87}. Specifically,
according to recent results \cite{88}, the modification (\ref{eq35}) made in \cite{86}
leads to a violation of the principle of gauge invariance.

\begin{figure}[b]
\vspace*{-11.cm}
\centerline{\hspace*{-1.5cm}
\includegraphics[width=7.in]{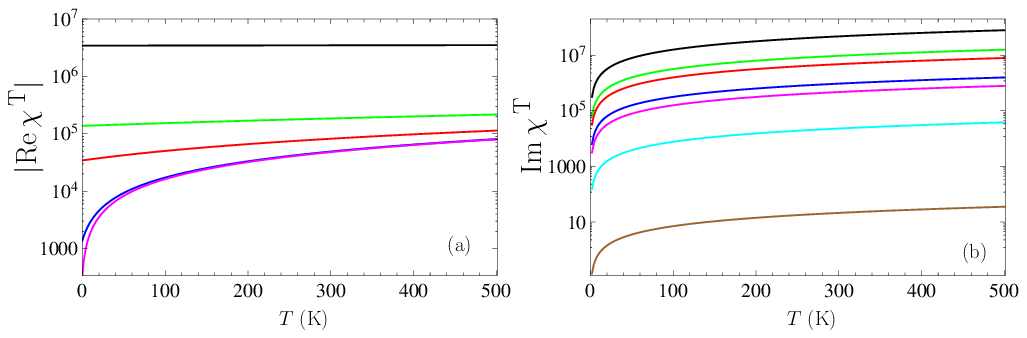}}
\vspace*{-10.2cm}
\caption{The computational results for (a) magnitude of the real part and (b)
imaginary part of the transverse electric susceptibility of graphene in the
region below the threshold are plotted as the functions of temperature.
The lines representing $|{\rm Re}\tx|$ in Figure~2(a) counted from top to
bottom are plotted for $\omega=10^7$, $5\times 10^7$, $10^8$, $5\times 10^8$,
and $10^9~$rad/s, respectively.
The lines representing ${\rm Im}\tx$ in Figure~2(b) counted from top to
bottom are plotted for $\omega=10^7$, $5\times 10^7$, $10^8$, $5\times 10^8$,
$10^9~$rad/s,
 $9\times10^9$, and $0.999999\times 10^{10}~$rad/s, respectively. }
\label{fg2}
\end{figure}
The computational results for the magnitude of real and imaginary parts of the transverse
electric susceptibility $\tx$ given by (\ref{eq29a}) and (\ref{eq30})
at $k=100~\mbox{cm}^{-1}$ are shown as the functions of temperature in
Figure~\ref{fg2}(a,b), respectively. The lines in Figure~2(a) counted from top to
bottom show the values of  $|{\rm Re}\tx|$ computed for different values of
$\omega=10^7$, $5\times 10^7$, $10^8$, $5\times 10^8$, and $10^9~$rad/s, respectively.
It is seen that at lower frequencies $|{\rm Re}\tx|$ is almost temperature-independent,
but at higher frequencies the dependence on $T$ becomes more pronounced.
By and large  $|{\rm Re}\tx|$ increases monotonously with increasing temperature
but decreases with increasing frequency.

In Figure~2(b), the lines counted from top to
bottom are plotted for ${\rm Im}\tx$ computed for the values of
$\omega=10^7$, $5\times 10^7$, $10^8$, $5\times 10^8$, $10^9$,
 $9\times10^9$, and $0.999999\times 10^{10}~$rad/s, respectively.
 At all frequencies, ${\rm Im}\tx$ increases monotonously with increasing
 temperature. With increasing frequency, ${\rm Im}\tx$ decreases
 at all temperatures.

\section{Temperature Dependence of the Dielectric Functions of Graphene
Above the Threshold}

We are coming now to the electric susceptibilities, $\lx$, $\tx$ and the
dielectric functions of graphene $\lve$, $\tve$ in the region above the
threshold $\omega>v_Fk$.

In this region, the real parts of the longitudinal electric susceptibility and
dielectric function are obtained from (\ref{eq9}) and (\ref{eq17})
\begin{eqnarray}
&&
{\rm Re}\lx\wkt={\rm Re}\lve\wkt-1=
\frac{2e^2}{v_F^2\hbar k}\left\{
\vphantom{\left[\int\limits_{0}^{v_Fk}\right]}
\frac{4k_BT \ln 2}{\hbar}\right.
\label{eq36}\\
&&~
-\left.\frac{1}{\wvk}\left[
\int\limits_{0}^{\infty}dxw(x,T)f_3(x)
-\int\limits_{v_Fk+\omega}^{\infty}\!\!\!dx w(x,T)f_4(x)+
\int\limits_{0}^{v_Fk-\omega}\!\!\!dx w(x,T)f_4(x)
\right]\right\}.
\nonumber
\end{eqnarray}

The imaginary parts of the longitudinal electric susceptibility and
dielectric function in the region $\omega>v_Fk$ are found
from (\ref{eq10}) and (\ref{eq17})
\begin{equation}
{\rm Im}\lx\wkt={\rm Im}\lve\wkt=
\frac{e^2}{2\hbar k\wvk}
\left[\pi k^2-\frac{4}{v_F^2}
\int\limits_{-v_Fk}^{v_Fk}\!\!\!dx w(\omega+x,T)
\sqrt{v_F^2k^2-x^2}\right].
\label{eq37}
\end{equation}

{}From (\ref{eq36}) and (\ref{eq37}) it can be seen that
\begin{equation}
\lim\limits_{\omega\to \infty}{\rm Re}\lx\wkt=
\lim\limits_{\omega\to \infty}{\rm Im}\lx\wkt= 0
\label{eq38}
\end{equation}
\noindent
as it should be. From (\ref{eq37}) it also follows that ${\rm Im}\lve>0$.

\begin{figure}[b]
\vspace*{-11.cm}
\centerline{\hspace*{-1.5cm}
\includegraphics[width=7in]{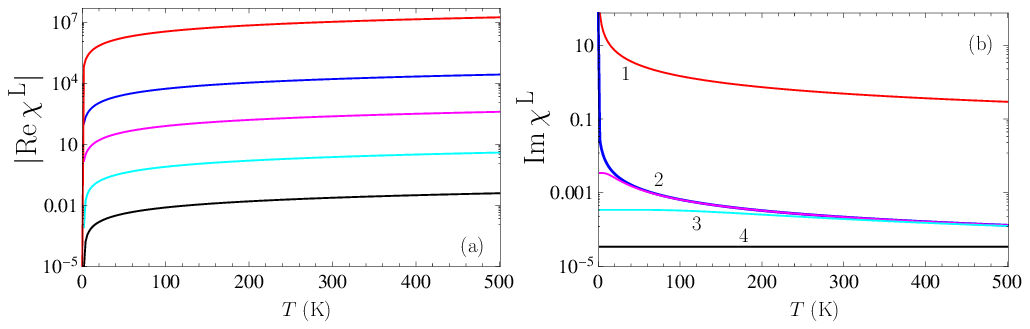}}
\vspace*{-10.2cm}
\caption{The computational results for (a) magnitude of the real part and (b)
imaginary part of the longitudinal electric susceptibility of graphene in the
region above the threshold are plotted as the functions of temperature.
The lines counted from top to bottom in Figure~3(a) are plotted for
$\omega=1.00001\times 10^{10}$, $1.5\times 10^{10}$, $10^{11}$,
$10^{12}$, and $10^{13}~$rad/s, respectively.
The lines in Figure~3(b) labeled 1, 2, 3, and 4
 are plotted for  $\omega=1.00001\times 10^{10}$, from $1.5\times 10^{10}$
 to $10^{11}$ (where the frequency-dependence is present only at low
frequencies), $10^{12}$, and  $10^{13}~$rad/s, respectively.}
\label{fg3}
\end{figure}

We have computed ${\rm Re}\lx$ and ${\rm Im}\lx$ by (\ref{eq36}) and (\ref{eq37})
taken at $k=100~\mbox{cm}^{-1}$ in the region above the threshold $\omega>v_Fk$ as
the functions of temperature. The computational results for $|{\rm Re}\lx|$ are
presented in Figure~\ref{fg3}(a) for the values of
$\omega=1.00001\times 10^{10}$, $1.5\times 10^{10}$, $10^{11}$,
$10^{12}$, and $10^{13}~$rad/s by the lines counted from top to bottom, respectively.
For ${\rm Im}\lx$, the computational results  are
shown in Figure~\ref{fg3}(b) for the values of
$\omega=1.00001\times 10^{10}$, from $1.5\times 10^{10}$ to $10^{11}$,
$10^{12}$, and $10^{13}~$rad/s by the respective lines labeled 1, 2, 3,
and 4. In doing so, line 2 corresponds to the frequency region
from $1.5\times 10^{10}$ to $10^{11}~$rad/s, where ${\rm Im}\lx$
does not depend on frequency with exception of only temperature interval
from 0 to 40~K. In this interval, line 2 is split into two such that
the upper one is for the frequency  $\omega=10^{11}~$rad/s and the
lower one for $\omega=1.5\times 10^{10}~$rad/s.

As is seen in Figure~\ref{fg3}(a,b), both $|{\rm Re}\lx|$ and ${\rm Im}\lx$
are the decreasing functions with increasing frequency. At the same time,
$|{\rm Re}\lx|$ increases monotonously with increasing temperature,
whereas ${\rm Im}\lx$ decreases with increasing temperature and for
sufficiently high frequencies becomes almost constant.

Next, we consider the real and imaginary parts of the transverse electric
susceptibility and dielectric function in the region $\omega>v_Fk$.
Thus, the real parts of these quantities are obtained by substituting
(\ref{eq11}) in (\ref{eq18})
\begin{eqnarray}
&&
{\rm Re}\tx\wkt={\rm Re}\tve\wkt-1=
-\frac{2e^2}{v_F^2\hbar k}\left\{
\vphantom{\left[\int\limits_{0}^{v_Fk}\right]}
\frac{4k_BT \ln 2}{\hbar}-\frac{\vkw}{\omega^2}
\right.
\label{eq39} \\
&&~
\times \left.
\left[
\int\limits_{0}^{\infty}dxw(x,T)\frac{(x+\omega)^2}{f_3(x)}
-\int\limits_{v_Fk+\omega}^{\infty}\!\!\!dx w(x,T)
\frac{(x-\omega)^2}{f_4(x)}+
\int\limits_{0}^{v_Fk-\omega}\!\!\!dx w(x,T)\frac{(x-\omega)^2}{f_4(x)}
\right]\right\}.
\nonumber
\end{eqnarray}
\noindent
The imaginary parts of $\tx$ and $\tve$ in the region $\omega>v_Fk$ are found
from the substitution (\ref{eq12}) in (\ref{eq18})
\begin{equation}
{\rm Im}\tx\wkt={\rm Im}\tve\wkt=
\frac{e^2\wvk}{2\hbar k\omega^2}\left[\pi k^2-
\frac{4}{v_F^2}\int\limits_{-v_Fk}^{v_Fk}\!\!\!dx w(\omega+x,T)
\frac{x^2}{\sqrt{v_F^2k^2-x^2}}
\right].
\label{eq40}
\end{equation}

Similar to the case of longitudinal quantities,
from (\ref{eq39}) and (\ref{eq40}) it follows that
\begin{equation}
\lim\limits_{\omega\to \infty}{\rm Re}\tx\wkt=
\lim\limits_{\omega\to \infty}{\rm Im}\tx\wkt= 0
\label{eq41}
\end{equation}
\noindent
and from (\ref{eq40}) it can be seen that ${\rm Im}\tve>0$.

\begin{figure}[h]
\vspace*{-10.5cm}
\centerline{\hspace*{-1.5cm}
\includegraphics[width=7.in]{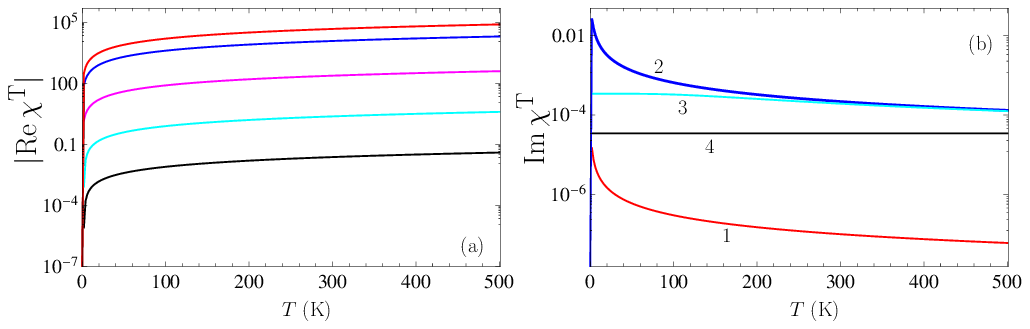}}
\vspace*{-10.2cm}
\caption{The computational results for (a) magnitude of the real part and (b)
imaginary part of the transverse electric susceptibility of graphene in the
region above the threshold are plotted as the functions of temperature.
The lines counted from top to bottom in Figure~4(a) are plotted for
$\omega=1.00001\times 10^{10}$, $1.5\times 10^{10}$, $10^{11}$,
$10^{12}$, and $10^{13}~$rad/s, respectively.
The lines in Figure~4(b) labeled 1, 2, 3, and 4
 are plotted for  $\omega=1.00001\times 10^{10}$, from $1.5\times 10^{10}$
 to $10^{11}$ (where the frequency dependence is lacking),
 $10^{12}$, and  $10^{13}~$rad/s, respectively.}
\label{fg4}
\end{figure}

In Figure~\ref{fg4}(a,b), the computational results for $|{\rm Re}\tx|$ and
${\rm Im}\tx$, respectively, at $k=100~\mbox{cm}^{-1}$ are presented as the
functions of temperature (a) by the lines counted from top to bottom
computed for $\omega=1.00001\times 10^{10}$, $1.5\times 10^{10}$, $10^{11}$,
$10^{12}$, and $10^{13}~$rad/s and (b) by the lines labeled 1, 2, 3, and 4
computed for $\omega=1.00001\times 10^{10}$, from $1.5\times 10^{10}$ to
$10^{11}$ (in this frequency region ${\rm Im}\tx$ does not depend on frequency),
$10^{12}$, and $10^{13}~$rad/s, respectively.

As is seen in Figur~\ref{fg4}(a),  $|{\rm Re}\tx|$ decreases monotonously with
increasing frequency. This is, however, not the case for  ${\rm Im}\tx$ which
depends on frequency nonmonotonously by increasing when $\omega$ changes
from $1.00001\times 10^{10}$ to $10^{11}~$rad/s and than decreasing with further
increase of $\omega$ to $10^{13}~$rad/s.

To conclude this section, we note that the polarization tensor and, as a consequence,
the response functions of graphene, are analytic in the upper plane of complex
frequencies. Because of this, both $\lve$ and $\tve$ satisfy the Kramers-Kronig
relations with the necessary number of subtractions. In so doing, there is no
subtraction for $\lve$ which is regular at zero frequency. The presence of a simple
pole in ${\rm Re}\tx$ and ${\rm Im}\tx$ results in one subtraction each (compare
with the dielectric function of usual metals where the single pole in the imaginary
part of the dielectric function results in the corresponding subtraction in the
Kramers-Kronig relation \cite{74}). One more subtraction in the  Kramers-Kronig
relation arises due to the presence of a double pole in ${\rm Re}\tx$.
The specific form of the resulting Kramers-Kronig relation  is considered
in \cite{89}.

\section{Thermal Effects in the Casimir Force Between Two Graphene Sheets}

The Casimir force per unit area of two parallel graphene sheets separated by
a distance $a$, i.e., the Casimir pressure, is given by the Lifshitz formula
\cite{5,6,7,8,9}
\begin{equation}
P(a,T)=-\frac{k_BT}{\pi}\sum_{l=0}^{\infty}\left(1-\frac{\delta_{l0}}{2}\right)
\int\limits_{0}^{\infty}q_lkdk\left\{
\left[r_{\rm TM}^{-2}(i\xi_l,k,T)\,e^{2aq_l}-1\right]^{-1}
+\left[r_{\rm TE}^{-2}(i\xi_l,k,T)\,e^{2aq_l}-1\right]^{-1}\right\},
\label{eq42}
\end{equation}
\noindent
where $\delta_{ln}$ is the Kronecker symbol,
$q_l\equiv q(i\xi_l)=(k^2+\xi_l^2/c^2)^{1/2}$, $\xi_l=2\pi k_BTl/\hbar$,
$l=0,\,1,\,2,\,\ldots$ are the Matsubara frequencies, and $r_{\rm TM}$ and
$r_{\rm TE}$ are the reflection coefficients on a graphene sheet.

In the literature, the Casimir pressure between two graphene sheets was calculated
in the framework of different theoretical approaches using various forms of response
functions of graphene to the electromagnetic field. Thus, these calculations were
performed using the hydrodynamic model of graphene \cite{18,19}, density-density
correlation functions \cite{45,91}, by modeling the response functions by means
of Lorentz-type oscillators \cite{44,46,92} etc.

As mentioned in Section~1, the most important breakthrough was reached in \cite{15}.
It lies in discovering the fact that the thermal regime in graphene systems starts at much
shorter separations than for the ordinary 3D bodies. As mentioned in Section~1,
this is partially explained
by the point that in addition to the standard effective temperature defined as
$k_BT_{\rm eff}=\hbar c/(2a)$, which arises from interaction with the electromagnetic
field, there is one more effective temperature for graphene
$k_BT_{\rm eff}^g=\hbar v_F/(2a)$ which is much lower.  Below we briefly review the
main characteristic features of the thermal effects in the Casimir force for
graphene systems using the most fundamental formalism of the polarization tensor.

To calculate the Casimir pressure (\ref{eq42})  using this formalism, it is necessary
to find the reflection coefficients at the pure imaginary Matsubara frequencies
$\omega=i\xi_l$. These are obtained using the expressions for $\lve$ and $\tve$
in (\ref{eq27}), (\ref{eq28}) and (\ref{eq29a}), (\ref{eq30}) derived in the
region of real frequencies $\omega<v_Fk$, i.e., below the threshold \cite{66}.

 In doing so it is necessary, first, to combine the real and imaginary parts
 of each dielectric function. For instance,  using (\ref{eq27}) and (\ref{eq28})
 one obtains
\begin{eqnarray}
&&
\lve\wkt={\rm Re}\lve\wkt+i{\rm Im}\lve\wkt=1+
\frac{\pi e^2k}{2\hbar\vkw}+\frac{8e^2k_BT\ln 2}{v_F^2\hbar^2k}
\nonumber \\
&&~~~~~
+\frac{2 e^2}{v_F^2\hbar k\vkw}
\left[\int\limits_{0}^{\infty}\!dx w(x,T)f_1(x)-
\int\limits_{0}^{\infty}\!dx w(x,T)f_2(x)
\right].
\label{eq43}
\end{eqnarray}

Substituting here $\omega=i\xi_l$ with the appropriately chosen branches
of the square roots \cite{66}, we find \cite{16}
\begin{eqnarray}
&&
\lve(i\xi_l,\mbox{\boldmath$k$},T)=1+
\frac{\pi e^2k}{2\hbar\sqrt{v_F^2k^2+\xi_l^2}}+\frac{8e^2k_BT\ln 2}{v_F^2\hbar^2k}
\nonumber \\
&&~~~~~
-\frac{4e^2}{v_F^2\hbar k\sqrt{v_F^2k^2+\xi_l^2}}
\int\limits_{0}^{\infty}\!dx w(x,T){\rm Re}\sqrt{v_F^2k^2-(x-i\xi_l)^2}.
\label{eq44}
\end{eqnarray}

In a similar way, using (\ref{eq29a}) and (\ref{eq30}), for the transverse dielectric
function of graphene at the pure imaginary Matsubara frequencies we obtain \cite{16}
\begin{eqnarray}
&&
\tve(i\xi_l,\mbox{\boldmath$k$},T)=1+
\frac{\pi e^2k}{2\hbar\xi_l^2}\sqrt{v_F^2k^2+\xi_l^2}-\frac{8e^2k_BT\ln 2}{v_F^2\hbar^2k}
\label{eq45} \\
&&~~~~~
+\frac{4e^2\sqrt{v_F^2k^2+\xi_l^2}}{v_F^2\hbar k\xi_l^2}
\int\limits_{0}^{\infty}\!dx w(x,T)\left[{\rm Re}\sqrt{v_F^2k^2-(x-i\xi_l)^2}
-{\rm Re}\frac{v_F^2k^2}{\sqrt{v_F^2k^2-(x-i\xi_l)^2}}\right].
\nonumber
\end{eqnarray}

Computations of the thermal Casimir pressure between two pristine graphene sheets by
equations equivalent to (\ref{eq42}), (\ref{eq23}), (\ref{eq24}), (\ref{eq44}), and
(\ref{eq45}) were performed in \cite{93}. It was shown that at separations from 10
to 20~nm the magnitudes of the Casimir pressure computed at $T=300~$K are far in excess
of those computed at $T=0~$K.  This confirmed the presence of unusually big thermal
effect in graphene systems which was observed experimentally later on \cite{84,85}.

The role of an explicit thermal effect due to a dependence of the polarization tensor
and the dielectric functions on temperature as a parameter was investigated in \cite{16}.
It was shown that at moderate separations the explicit thermal effect in the Casimir
pressure contributes to the total thermal correction nearly equally to the implicit
thermal effect originating from a summation over the Matsubara frequencies.

\begin{figure}[h]
\vspace*{-10.5cm}
\centerline{\hspace*{-1.5cm}
\includegraphics[width=7.in]{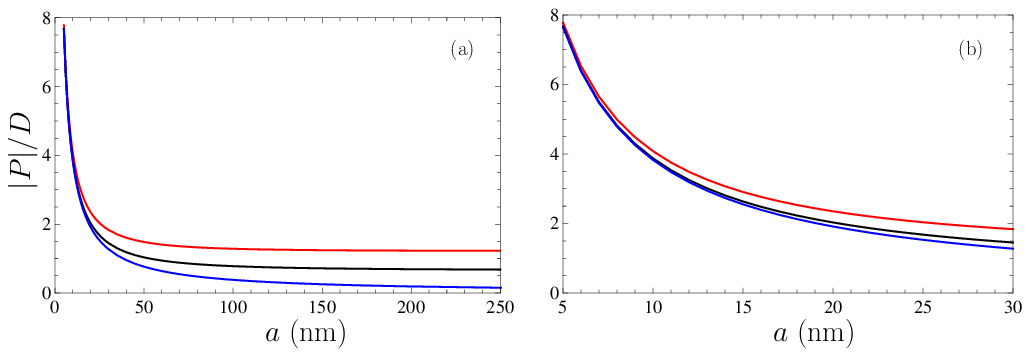}}
\vspace*{-10.2cm}
\caption{The computational results for the normalized magnitude of the Casimir
pressure between two graphene sheets are shown as the function of separation
by the upper, medium and bottom lines computed at $T=300~$K exactly, at $T=300~$K with
taking into account only an implicit thermal effect, and at $T=0~$K, respectively,
over the separation region (a) from 2 to 250~nm and (b) from 5 to 30~nm.}
\label{fg5}
\end{figure}

This result is illustrated in Figure~\ref{fg5}(a,b) where the magnitude of the Casimir
pressure normalized to the quantity $D=k_BT/(8\pi a^3)$ is shown as the function of
separation by the three lines, where the top and medium lines are computed at
$T=300~$K exactly and taking into account only an implicit temperature dependence,
respectively, whereas the bottom line is computed at $T=0~$K. In Figure~\ref{fg5}(b),
the separation region from 5 to 30~nm is shown on an enlarged scale for better
visualization.

In Fig.~\ref{fg5}(a,b), the big total thermal effect is characterized by the
difference between the top and bottom lines. It consists of two parts.
The first of them is a difference between the intermediate and bottom lines.
It is an implicit contribution due to a summation over the Matsubara frequencies.
The second part is a difference between the top and intermediate lines which is
an explicit thermal effect caused by a dependence of the response functions of
graphene on temperature.

In the high temperature (large separations) limit, the Casimir pressure between
two graphene sheets admits an analytic representation \cite{93}
\begin{equation}
P(a,T)=-\frac{k_BT\zeta(3)}{8\pi a^3}\left(1-
\frac{3v_F^2\hbar^2}{8\ln 2 e^2ak_BT}\right),
\label{eq46}
\end{equation}
\noindent
where $\zeta(z)$ is the Riemannian zeta function. At $T=300~$K, the Casimir pressure
calculated by (\ref{eq46}) agrees with the results of numerical computations to
within 1\% at all separations exceeding 370~nm. Already at separation of 200~nm,
the first, classical, term in (\ref{eq46}) contributes 96.9\% of the thermal
Casimir pressure.

Impact of the nonzero mass gap $\Delta$ and chemical potential $\mu$ of graphene
sheets on the thermal Casimir force acting between them was investigated in
\cite{93,94}. It was shown that for $\Delta\neq 0$ the Casimir pressure remains
constant with increasing temperature within some temperature interval.
This temperature interval is wider for larger $\Delta$. Thus, if $\Delta\neq 0$,
the thermal effect in the Casimir interaction between graphene sheets is suppressed.
The nonzero chemical potential $\mu$ acts on the thermal Casimir pressure in the
opposite direction. By and large the magnitude of the Casimir pressure increases with
increasing $\mu$ and decreases with increasing $\Delta$.  {Using the formalism
of the polarization tensor, the thermal Casimir force in the system of $N$ parallel
2${\rm D}$ Dirac materials was considered in} \cite{94a}.

In experiments, graphene sheets are usually deposited on some substrates.
In this case one should use the Lifshitz formula (\ref{eq42}) where the reflection
coefficients $r_{\rm TM,TE}$ defined in (\ref{eq23}) and (\ref{eq24}) are
replaced with $R_{\rm TM,TE}$ defined in (\ref{eq25}) and (\ref{eq26}).
The thermal Casimir interaction between two graphene-coated plates was investigated
in \cite{95}. It was shown that the Casimir pressure between two metallic plates
is almost unaffected by the graphene coatings. If, however, the substrates are made of a dielectric
material (fused silica glass, SiO$_2$, for instance), the presence of graphene
coatings significantly increases the magnitudes of the total Casimir pressure.
As to the magnitude of the thermal correction and its fractional weight in the total
Casimir pressure, both are smaller than for the freestanding graphene sheets \cite{94}.
It was also shown that for the graphene-coated plates the influence of nonzero $\Delta$
and $\mu$ on the Casimir pressure is much smaller than for the freestanding graphene
sheets, although the qualitative character of their impact remains the same \cite{94}.

Note also that an investigation of the thermal Casimir interaction between a
freestanding graphene sheet and either a metallic or a dielectric
plate was performed in \cite{96}. In this case, the factors $r_{\rm TM,TE}^{-2}$
in (\ref{eq42}) are replaced with $r_{\rm TM,TE}^{-1}\tilde{r}_{\rm TM,TE}^{-1}$
where $\tilde{r}_{\rm TM,TE}$ are the standard Fresnel reflection coefficients
on a material plate defined as
\begin{equation}
\tilde{r}_{\rm TM}(i\xi_l,\mbox{\boldmath$k$})=
\frac{\ve(i\xi_l)q-\tilde{q}}{\ve(i\xi_l)q+\tilde{q}},
\qquad
\tilde{r}_{\rm TE}(i\xi_l,\mbox{\boldmath$k$})=
\frac{q-\tilde{q}}{q+\tilde{q}}.
\label{eq47}
\end{equation}

It was shown \cite{96} that for a pristine graphene sheet the thermal correction
remains rather large as compared with the case of two plates made of the
ordinary 3D materials. For a graphene sheet with a relatively large $\Delta$,
the thermal correction remains negligibly small within some temperature interval.

\section{Thermal Effects in the Casimir-Polder Force Between
a Nanoparticle and a Graphene Sheet}

The Casimir-Polder force between a small particle spaced at a height $a$ above
a graphene sheet is given by the Lifshitz formula \cite{4,5,9}
\begin{equation}
F(a,T)=-\frac{2k_BT}{c^2}\sum_{l=0}^{\infty}\left(1-\frac{\delta_{l0}}{2}\right)
\alpha(i\xi_l)\int\limits_{0}^{\infty}\!\!kdke^{-2aq_l}
\left[(2k^2c^2+\xi_l^2)r_{\rm TM}(i\xi_l,k,T)-
\xi_l^2r_{\rm TE}(i\xi_l,k,T)\right],
\label{eq48}
\end{equation}
\noindent
where $\alpha(i\xi_l)$ is the dynamic electric polarizability of a particle
calculated at the pure imaginary Matsubara frequencies and the reflection
coefficients on a graphene sheet are defined in (\ref{eq23}) and (\ref{eq24}).

Similar to the Casimir force between two graphene sheets, the Casimir-Polder
force with graphene was calculated in the literature using different theoretical
formalisms \cite{97,98,99,100,101,102,103,103a}. Computations of this force by
equation (\ref{eq48}) with the reflection coefficients equivalent to
(\ref{eq23}), (\ref{eq24}) and dielectric functions (\ref{eq44}), (\ref{eq45})
were performed in \cite{104,105}. It was shown that, similar to the case of
two parallel graphene sheets, there is big thermal effect in the Casimir-Polder
force already at relatively short separations.
\begin{figure}[b]
\vspace*{-21.cm}
\centerline{\hspace*{-1.5cm}
\includegraphics[width=15.in]{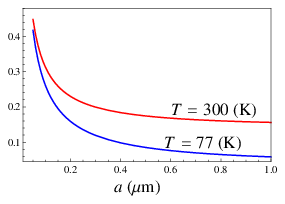}}
\vspace*{-27.cm}
\caption{The computational results for the magnitude of the Casimir-Polder force
between an atom of metastable helium and a graphene sheet multiplied by the factor
$a^4$ are shown as a function of separation.
The top line is computed at $T=300~$K and the bottom line at $T=77~$K.}
\label{fg6}
\end{figure}
To illustrate this result, in Figure~\ref{fg6} we plot the magnitude of the
Casimir-Polder force between an atom of metastable helium, He$^\ast$, and
a graphene sheet multiplied by the factor $a^4$ as a function of atom-graphene
separation. The top line is computed at $T=300~$K and the bottom line at $T=77~$K
\cite{105}. As is seen in Figure~\ref{fg6}, the magnitude of the Casimir-Polder
force increases significantly with increasing temperature already at separations
of 100--200~nm.

In the limit of high temperatures (large separations) the Casimir-Polder force
can be expressed analytically \cite{106}
\begin{equation}
F(a,T)=-\frac{3k_BT}{4 a^4}\alpha(0)\left(1-
\frac{\hbar^2v_F^2}{4\ln 2 e^2k_BTa}\right).
\label{eq49}
\end{equation}
\noindent
This expression gives more than 98\% of the total Casimir-Polder force at
separations exceeding $1.5~\mu$m. Thus, the Casimir-Polder force from graphene
reaches its asymptotic regime at larger separations than the Casimir force
between two graphene sheets (see Section 5), but at by a factor of 4 shorter
separations than in the case of ordinary materials \cite{5}.

Similar to the case of two graphene sheets, the nonzero mass gap and chemical
potential of a graphene sheet act on the Casimir-Polder force in the opposite
directions by decreasing and increasing its magnitude, respectively \cite{105}.
The asymptotic expression of large separations with account of nonzero
$\Delta$ and $\mu$ was obtained in \cite{105a}.

For calculation of the Casimir-Polder force between a nanoparticle and a
graphene-coated substrate, the reflection coefficients $r_{\rm TM,TE}$ in
(\ref{eq48}) should be replaced with $R_{\rm TM,TE}$ defined  in
(\ref{eq25}) and (\ref{eq26}). Computations performed for a He$^\ast$ atoms
above a graphene-coated SiO$_2$ substrate show that the presence of a graphene
coating increases the magnitude of the Casimir-Polder force \cite{105}.
For a substrate coated with gapped and doped graphene, the magnitude of
the Casimir-Polder force decreases with increasing $\Delta$ and increases
with increasing $\mu$. These effects have a simple physical explanation.
The point is that an increase of $\Delta$ results in a decreased mobility
of charge carriers and, thus, in decreased conductivity of graphene.
Just to the opposite, an increase of $\mu$ leads to a larger density of
charge carriers and, thus, to a larger conductivity.
The asymptotic expression of large separations for a gapped and doped graphene
sheet deposited on a substrate was  found in \cite{105b}.

\section{Thermal Effects in the Casimir and Casimir-Polder Forces from
Graphene Out of Thermal Equilibrium}

The Lifshitz formulas for the Casimir (\ref{eq42}) and Casimir-Polder
(\ref{eq48}) forces were derived \cite{7,8,9} for systems in the state
of thermal equilibrium, i.e., under a condition that temperature of all
interacting bodies is the same as that of the environment.
If the temperature of at least one body is different from the environmental
temperature, the condition of thermal equilibrium is violated. Keeping in
mind, however, that the correlations of the polarization field expressed
by the fluctuation-dissipation theorem are spatially local, it is natural
to assume that in out-of-thermal-equilibrium situation they are given by
the same expressions but with appropriate temperatures \cite{107}.
This is a condition of the so-called local thermal equilibrium.

Under the condition of local thermal equilibrium, the Lifshitz theory of
the Casimir force was generalized for out-of-thermal-equilibrium situations
\cite{108,109,110,111}. The created formalism was then adapted for the
case of arbitrary shaped bodies kept at different temperatures
\cite{112,113,114,115,116,117,118} and possessing the temperature-dependent
dielectric functions \cite{119,120} like this is the case for graphene.

Here, we present an expression for the nonequilibrium Casimir pressure on
the lower graphene sheet where the upper one is kept at the environmental
temperature $T$ and the lower one has a different temperature $T_1$.
According to \cite{111}, this force can be conveniently presented as a sum
of two contributions
\begin{equation}
P_{\rm neq}(a,T,T_1)=\widetilde{P}_{\rm eq}(a,T,T_1)+
\Delta P_{\rm neq}(a,T,T_1),
\label{eq50}
\end{equation}
\noindent
where $\widetilde{P}_{\rm eq}$  is the mean of quasi-equilibrium contributions
taken at temperatures $T$ and $T_1$
\begin{equation}
\widetilde{P}_{\rm eq}(a,T,T_1)=\frac{1}{2}\left[
P(a,T;T_1)+P(a,T_1;T)\right].
\label{eq51}
\end{equation}
\noindent
Note that the first temperature argument in $P(a,T;T_1)$ indicates the temperature,
at which the Matsubara frequencies are calculated, whereas the second is the
temperature of graphene sheet different from that of the Matsubara frequencies
(i.e., $T_1$ in the first case and $T$ in the second).
Using (\ref{eq42}), we represent (\ref{eq51}) in the form
\begin{eqnarray}
&&
\widetilde{P}_{\rm eq}(a,T,T_1)=-\frac{k_B}{2\pi}\sum_{l=0}^{\infty}
\left(1-\frac{\delta_{l0}}{2}\right)\left\{\!
T\!\!\int\limits_{0}^{\infty}\!\!q_lkdk\sum_{\kappa}\left[
r_{\kappa}^{-1}(i\xi_l,k,T)r_{\kappa}^{-1}(i\xi_l,k,T_1)\,e^{2aq_l}-1
\right]^{-1}
\right.
\nonumber\\
&&~~~~~~~~~~
\left.
+T_1\int\limits_{0}^{\infty}\!\!q_l^{(1)}kdk\sum_{\kappa}\left[
r_{\kappa}^{-1}(i\xi_l^{(1)},k,T)r_{\kappa}^{-1}(i\xi_l^{(1)},k,T_1)\,
e^{2aq_l^{(1)}}-1\right]^{-1}
\right\}.
\label{eq52}
\end{eqnarray}
\noindent
Here, the sum in $\kappa$ is over two polarizations of the electromagnetic
field, $\kappa={\rm TM,\,TE}$, $\xi_l^{(1)}=2\pi k_BT_1l/\hbar$,
$q_l^{(1)}=(k^2+{\xi_l^{(1)}}^2/c^2)^{1/2}$ and the reflection
coefficients on a graphene sheet are defined in (\ref{eq23}) and (\ref{eq24}).

The second term on the r.h.s. of (\ref{eq50}) is the proper nonequilibrium
contribution given by \cite{111,114}
\begin{eqnarray}
&&
\Delta P_{\rm neq}(a,T,T_1)=\frac{\hbar}{4\pi^2}\int\limits_{0}^{\infty}\!d\omega
\left[\Theta(\omega,T)-\Theta(\omega,T_1)\right]
\int\limits_{0}^{\omega/c}\!pkdk\sum_{\kappa}
\frac{|r_{\kappa}(\omega,k,T_1)|^2-
|r_{\kappa}(\omega,k,T)|^2}{|B_{\kappa}(\omega,k,T,T_1)|^2}
\label{eq53}\\
&&
-\frac{\hbar}{2\pi^2}\int\limits_{0}^{\infty}\!\!d\omega
\left[\Theta(\omega,T)-\Theta(\omega,T_1)\right]
\int\limits_{\omega/c}^{\infty}\!\!k{\rm Im}pdk
e^{-2a{\rm Im p}}
\sum_{\kappa}
\frac{{\rm Im}r_{\kappa}(\omega,k,T)
{\rm Re}r_{\kappa}(\omega,k,T_1)-{\rm Re}r_{\kappa}(\omega,k,T)
{\rm Im}r_{\kappa}(\omega,k,T_1)}{|B_{\kappa}(\omega,k,T,T_1)|^2},
\nonumber
\end{eqnarray}
\noindent
where
\begin{equation}
\Theta(\omega,T)=\left[\exp\left(\frac{\hbar\omega}{k_BT}\right)-1\right]^{-1},
\qquad
p=\sqrt{\frac{\omega^2}{c^2}-k^2}
\label{eq54}
\end{equation}
\noindent
and
\begin{equation}
B_{\kappa}(\omega,k,T,T_1)=1- r_{\kappa}(\omega,k,T)r_{\kappa}(\omega,k,T_1)
\,e^{2ipa}.
\label{eq55}
\end{equation}
\noindent
Note that both the propagating waves with $k\leqslant\omega/c$ and the
evanescent ones with $k>\omega/c$ contribute to (\ref{eq53}).

Thus, to compute the total nonequilibrium Casimir pressure on a lower graphene
sheet (\ref{eq50}), it is necessary to use the response functions of graphene
along the imaginary frequency axis for computations of the quasi-equilibrium
contribution (\ref{eq52}) and along the real frequency axis for computation of
the proper nonequilibrium contribution (\ref{eq53}). Computations of this kind
were performed in \cite{121,122}. It was shown that for a hotter and colder
graphene sheets than the environment the effects of nonequilibrium increase
and decrease the magnitude of the equilibrium Casimir pressure, respectively.

Computations of the noneuilibrium Casimir force were also performed for the case
of graphene-coated SiO$_2$ plates. For this purpose, the reflection coefficients
$r_{\rm TM,TE}$ in (\ref{eq52}) and (\ref{eq53}) should be replaced with
$R_{\rm TM,TE}$ defined in (\ref{eq25}) and (\ref{eq26}).
The computational results show that the presence of graphene coating leads to
an increased magnitude of the nonequilibrium Casimir force. For higher temperature
and chemical potential of a graphene coating, this increase is greater as well
as for smaller energy gap.

The generalization of the Lifshitz theory for out-of-thermal-equilibrium situations
makes it possible to calculate the nonequilibrium Casimir-Polder force acting
between an atom or a nanoparticle and  a graphene sheet. This generalization was
performed in \cite{123,124}.

We consider a spherical nanoparticle of radius $R$ kept at the environmental
temperature $T$ at the height $a$ above a graphene sheet kept at temperature $T_1$
which can be either lower of higher than $T$. It is assumed that $R\ll a$,
$R\ll\hbar c/(k_BT)$, and $R\ll\hbar c/(k_BT_1)$ \cite{117}.
Recall that at $T=300~$K it holds
$\hbar c/(k_BT)\approx 7.6~\mu$m. Under these conditions it is possible to use
the static electric polarizability of a nanoparticle
\begin{equation}
\alpha_0=R^3\frac{\ve(0)-1}{\ve(0)+2}, \qquad
\alpha_0=R^3
\label{eq56}
\end{equation}
\noindent
for the dielectric and metallic nanoparticles, respectively.

Similar to the nonequilibrium Casimir pressure (\ref{eq50}), the nonequilibrium
Casimir-Polder force can be presentes as a sum of the quasi-equilibrium and proper
nonequilibrium contributions. Here we use the representation \cite{70,113}
\begin{equation}
F_{\rm neq}(a,T,T_1)=\widetilde{F}_{\rm eq}(a,T;T_1)+
\Delta F_{\rm neq}(a,T;T_1),
\label{eq57}
\end{equation}
\noindent
where
\begin{equation}
\widetilde{F}_{\rm eq}(a,T;T_1)=-\frac{2k_BT\alpha_0}{c^2}
\sum_{l=0}^{\infty}\left(1-\frac{\delta_{l0}}{2}\right)
\int\limits_{0}^{\infty}\!\!kdke^{-2aq_l}
\left[(2k^2c^2+\xi_l^2)r_{\rm TM}(i\xi_l,k,T_1)-
\xi_l^2r_{\rm TE}(i\xi_l,k,T_1)\right],
\label{eq58}
\end{equation}
\noindent
and
\begin{equation}
\Delta F_{\rm neq}(a,T;T_1)=\frac{2\hbar\alpha_0}{\pi c^2}
\int\limits_{0}^{\infty}\!d\omega\Theta(\omega,T,T_1)
\int\limits_{\omega/c}^{\infty}\!\!kdke^{-2a{\rm Im}p}
{\rm Im}\left[(2k^2c^2-\omega^2)r_{\rm TM}(\omega,k,T_1)+
\omega^2r_{\rm TE}(\omega,k,T_1)\right].
\label{eq59}
\end{equation}
\noindent
Note that (\ref{eq58}) differs from the usual equilibrium Casimir-Polder force
(\ref{eq48}) by the temperature argument $T_1$ in the reflection coefficients,
whereas the Matsubara frequencies $\xi_l$ are calculated at the environmental
temperature $T$. The specific feature of (\ref{eq59}), as compared to (\ref{eq53}),
is that $\Delta F_{\rm neq}$ is determined by only the contribution of the
evanescent waves with $k>\omega/c$.

Computations of the nonequilibrium Casimir-Polder force between a spherical
nanoparticle and a pristine graphene sheet using (\ref{eq57})--(\ref{eq59}) and
(\ref{eq23}), (\ref{eq24}) were parformed in \cite{70}.
 Similar to the case of the nonequilibrium Casimir force between two graphene
 sheets, it was shown that the nonequilibrium effects increase the magnitude
 of the Casimir-Polder force for a hotter graphene sheet than the environment
 and decrease it for a cooler graphene sheet. Thus, in the case $T_1<T$,
 the nonequilibrium Casimir-Polder force may change its sign at some separation
 distance and become repulsive at larger separations.

\begin{figure}[t]
\vspace*{-5.5cm}
\centerline{\hspace*{-2.cm}
\includegraphics[width=7.in]{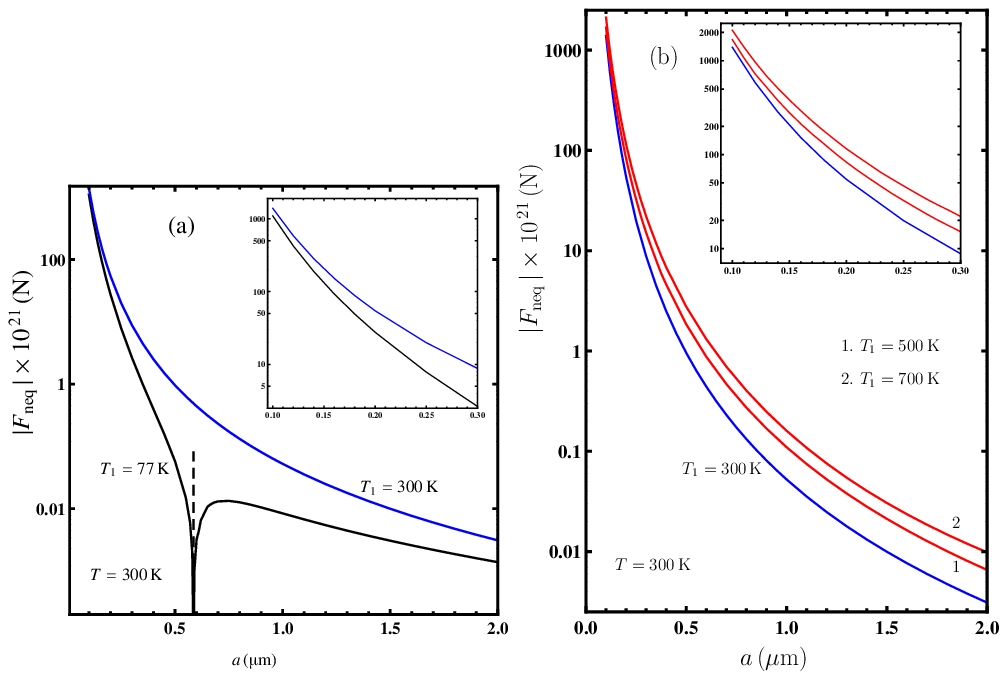}}
\vspace*{-10.cm}
\caption{The computational results for the magnitude of nonequilibrium
Casimir-Polder force between a metallic nanoparticle of 5~nm diameter and
(a) cooled down to $T_1=77~$K and (b) heated up to $T_1=500$ and 700~K
 multiplied by the factor
$10^{21}$ are shown as the functions of separation.
For comparison, (a) the top line and (b) the bottom line show the
equilibrium Casimir-Polder force computed at $T_1=T=300~$K.}
\label{fg7}
\end{figure}

 In Figure~\ref{fg7}(a,b) we plot the magnitude of the nonequilibrium
 Casimir-Polder force multiplied by the factor $10^{21}$ between a metallic
 nanoparticle of 5~nm diameter and either cooled down to $T_1=77~$K or heated
 up to $T_1=500$ and 700~K graphene sheet.

\noindent
For comparison purposes, the top line in Fig.~\ref{fg7}(a) and the bottom
line in Fig.~\ref{fg7}(b) are computed at the environmental temperature $T=300~$K.
The lines labeled 1 and 2 in Fig.~\ref{fg7}(b) are computed at $T_1=500$ and
700~K, respectively. On the insets, the regions of short graphene-nanoparticle
separations are shown on an enlarged scale.

{}From Figure~\ref{fg7}(a) it is seen that for a cooled graphene sheet the
Casimir-Polder force turns into zero at $a=0.58~\mu$m and becomes repulsive
at larger separations. All forces in Figure~\ref{fg7}(b), plotted for a heated
graphene sheets, are negative, i.e., attractive. With increasing temperature
from 500 to 700~K, the magnitude of the nonequilibrium Casimir-Polder force
increases.

The influence of the nonzero energy gap parameter of a graphene sheet on the
nonequilibrium Casimir-Polder force was investigated in \cite{125}. It was
shown that for a gapped graphene sheet the nonequilibrium Casimir-Polder force
preserves its sign even if it is cooled to lower temperatures than the
environmental one.  The impact of a substrate underlying the gapped graphene
sheet on the nonequilibrium Casimir-Polder force was analyzed in \cite{126}.
According to the results obtained, the presence of a substrate results in an
increased magnitude of the nonequilibrium Casimir-Polder force. However, with
increasing energy gap, the nonequilibrium Casimir-Polder force becomes smaller
and the impact of the graphene coating on the total force decreases.

The combined effect of the nonzero mass gap and chemical potential of graphene
coating on the nonequilibrium Casimir-Polder force was investigated in \cite{127}.
It was shown that with increasing $\mu$ the magnitude of the
nonequilibrium Casimir-Polder force increases irrespective of weather the
graphene-coated plate was heated or cooled. This increase is more pronounced when
the graphene-coated plate is cooled and less pronounced when it is heated.
The nonequilibrium Casimir-Polder force from a graphene-coated substrate is an
increasing function of temperature. The impact of the energy gap parameter
$\Delta$ on the Casimir-Polder force for a cooled graphene-coated plate with
nonzero $\mu$ is larger that for a heated one. With increasing separation between
a nanoparticle and a graphene-coated plate, the impact of temperature on the
nonequilibrium Casimir-Polder force from a graphene-coated plate becomes stronger.

\section{Discussion}
In the foregoing, we have considered the response functions of graphene which depend
not only of frequency but also on wave vector and temperature. It has been known
that the response functions of conventional materials, such as dielectrics, metals,
and semiconductors, are found using some phenomenological and partially
phenomenological approaches, such as Boltzmann's transport theory, Kubo's model,
the random phase approximation etc. In this regard, the novel two-dimensional
material graphene is unique because under the application conditions of the Dirac
model it is described by the relativistic thermal quantum field theory in
(2+1)-dimensional space-time. As a result, the response functions of graphene can be
found precisely starting from first physical principles and used for theoretical
description of various physical phenomena, such as the Casimir and Casimir-Polder
forces, radiative heat transfer, the conductivity and reflectivity properties of
graphene, etc.

As discussed above, all these effects are actively investigated using various
theoretical approaches. However, in the application region of the Dirac model,
i.e., at the characteristic energies below approximately 3~eV, where graphene can
be considered as a set of massless or very light free electronic quasiparticles
governed by the Dirac equation, the quantum field theoretical formalism using the
relativistic polarization tensor can be considered as a touchstone for all other
approaches.

In this regard, the prediction of the second order pole at zero frequency in the
transverse dielectric function of graphene made in the framework of quantum field
theoretical approach using the polarization tensor holds the greatest interest
today. The formalism incorporating this pole was used for a theoretical description
of the experimental data on measuring big thermal effect in graphene systems at
short separations and demonstrated a very good agreement with the measurement
data \cite{84,85}. The quantum field theoretical formalism using the polarization
tensor of graphene was also employed for the investigation of thermal dispersion
interaction of different atoms with graphene \cite{128,129,130,131,132},
calculation of the role of the uniaxial strain in the graphene
sheet \cite{132.1,132.2,132.3,132.4,132.5,132.6,132.7,132.9,132.10} and
in many other applications. It may cause further progress in studying the near-field
radiative heat transfer in graphene
systems \cite{132.11,132.12,132.13,132.14,132.15,132.16,132.17,132.18,132.19,132.20}.

Note that the second order pole in the transverse response function is not predicted
within the phenomenological and semi-phenomenological approaches, including the
Kubo model. Based on this, the attempt was undertaken \cite{86} mentioned in
Section~3 to consider it as "nonphysical". This conclusion is, however, scientifically
unwarranted because the theoretical approach starting from the fundamental physical
principles offers few advantages over the phenomenological and semi-phenomenological
ones. As always in physics, the last word in this discussion belongs to the
experiment.

As an exceptional novel material with outstanding electrical, optical, and mechanical
properties, graphene finds prospective applications in nanoelectronics
\cite{133,134,135,136,137}. At short separations characteristic for nanodevices both
the Casimir and Casimir-Polder forces take on great significance. This is the reason
why the reliable calculation methods of these forces discussed above are much needed
for further progress in the field.

\section{Conclusions}

Here, we investigated the temperature dependence of the spatially nonlocal response
functions of graphene expressed via the polarization tensor and reviewed their
applications to calculation of the Casimir and Casimir-Polder forces in and out of
thermal equilibrium. Simple and convenient in applications expressions for the real
and imaginary parts of the polarization tensor of a pristine graphene are presented.
This made it possible to analyze the temperature dependence of its longitudinal and
transverse response functions in the regions below and above the threshold.
The response functions of graphene satisfy the Kramers-Kronig relations and possess
all other necessary properties such as the positive imaginary part and approaching
unity in the limit of infinitely increasing frequency. The unusual novel property
is the presence of a double pole in the transverse response function which already
found an implicit confirmation in experiments on measuring  an unusually big thermal
effect in the Casimir force from graphene at short separations.

The thermal properties of the response functions of graphene were illustrated by
their impact on the Casimir and Casimir-Polder forces in and out of thermal equilibrium.
Thus, we considered the thermal effect in the equilibrium Casimir pressure between two
parallel graphene sheets and the Casimir-Polder force between an atom of metastable
helium and a graphene sheet. The relative roles of the implicit thermal effect arising
due to a summation over the Matsubara frequencies and the explicit one due to a
temperature dependence of the response functions of graphene were elucidated.
We concluded by considering the out-of-thermal-equilibrium Casimir force between
two graphene sheets and the Casimir-Polder force between a nanoparticle and graphene.
In all cases, the role of nonzero energy gap and chemical potential was specified, as
well as an impact on the force of a material substrate underlying the graphene sheet.

The presented formalism gives the possibility to reliably calculate the Casimir and
Casimir-Polder forces from graphene in and out of thermal equilibrium for applications
in both fundamental physics and nanotechnology.

\section*{Acknowledgments}
The authors are grateful to M. Bordag and N. Khusnutdinov
for useful discussions. This work was
supported by the State Assignment for Basic Research (project FSEG-2023-0016).


\end{document}